# DPIS: An Enhanced Mechanism for Differentially Private SGD with Importance Sampling

Jianxin Wei
National University of Singapore
Singapore
jianxinwei@comp.nus.edu.sg

Ergute Bao
National University of Singapore
Singapore
ergute@comp.nus.edu.sg

Xiaokui Xiao
National University of Singapore
Singapore
xkxiao@nus.edu.sg

Yin Yang
Hamad Bin Khalifa University
Qatar
yyang@hbku.edu.qa

## ABSTRACT

Nowadays, differential privacy (DP) has become a well-accepted standard for privacy protection, and deep neural networks (DNN) have been immensely successful in machine learning. The combination of these two techniques, *i.e.*, deep learning with differential privacy, promises the privacy-preserving release of high-utility models trained with sensitive data such as medical records. A classic mechanism for this purpose is DP-SGD, which is a differentially private version of the stochastic gradient descent (SGD) optimizer commonly used for DNN training. Subsequent approaches have improved various aspects of the model training process, including noise decay schedule, model architecture, feature engineering, and hyperparameter tuning. However, the core mechanism for enforcing DP in the SGD optimizer remains unchanged ever since the original DP-SGD algorithm, which has increasingly become a fundamental barrier limiting the performance of DP-compliant machine learning solutions.

Motivated by this, we propose DPIS, a novel mechanism for differentially private SGD training that can be used as a drop-in replacement of the core optimizer of DP-SGD, with consistent and significant accuracy gains over the latter. The main idea is to employ importance sampling (IS) in each SGD iteration for mini-batch selection, which reduces both sampling variance and the amount of random noise injected to the gradients that is required to satisfy DP. Although SGD with IS in the non-private setting has been well-studied in the machine learning literature, integrating IS into the complex mathematical machinery of DP-SGD is highly non-trivial; further, IS involves additional private data release which must be protected under differential privacy, as well as computationally intensive gradient computations. DPIS addresses these challenges through novel mechanism designs, fine-grained privacy analysis, efficiency enhancements, and an adaptive gradient clipping optimization. Extensive experiments on four benchmark datasets, namely MNIST, FMNIST, CIFAR-10 and IMDb, involving both convolutional and recurrent neural networks, demonstrate the superior effectiveness of DPIS over existing solutions for deep learning with differential privacy.

## 1 INTRODUCTION

Deep learning [20] is an effective machine learning technique that trains a neural network model by iteratively optimizing a target loss function over an input dataset, usually with the stochastic gradient descent (SGD) algorithm or its variants. The resulting model, consisting of weights learned from the data, may unintentionally memorize information about individual records in the training dataset. This is a serious concern especially when the data is sensitive, such as medical records, in which case an adversary may derive private information from the model weights, possibly in unforeseen ways. In the literature, various types of privacy threats have been identified in recent years, including membership inference attacks [31, 36, 48, 49, 63], model inference attacks [19, 60, 63], and extraction attacks [8, 53]. In addition, even innocuous-looking personal data could be used against the individual: for instance, a person's face images could be used to synthesize realistic videos [51, 65]. These threats make it difficult for researchers to share trained deep learning models..

Differential privacy (DP) is a strong and rigorous privacy protection scheme that has been widely implemented and deployed in industry, *e.g.*, by Apple [1], Google [2], and Uber [38]. Enforcing differential privacy in the model training process promises to prevent privacy violations while retaining the high utility of the deep learning model. However, this is a highly challenging task, as straightforward techniques for ensuring DP lead to overwhelming noise in the gradients during SGD training, which, in turn, leads to poor utility of the resulting model [14]. A breakthrough, namely DP-SGD [2], came in 2016, which achieves much-improved utility compared to previous methods, for publishing deep learning models under $(\epsilon, \delta)$-DP, elaborated in Section 2.

Specifically, DP-SGD injects random noise into the gradients computed in each iteration of the SGD algorithm. The noise is carefully calibrated through a sophisticated, fine-grained privacy loss analysis called moments accountant (MA). Subsequent work [33] further improves MA with a more precise privacy loss calculation based on the concept of Rényi differential privacy (RDP) [32]. As we explain later on Section 2, any RDP mechanism can be converted in a straightforward manner to a mechanism for enforcing $(\epsilon, \delta)$-DP. In the following, we use the term DP-SGD to refer to the improved version of the algorithm that injects noise guided by the privacy accounting method in [33], which has been implemented in the production-ready Tensorflow Privacy library[1].

Due to the noise injected into the gradients, there is still a considerable performance gap between a model trained with DP-SGD under typical privacy parameter settings, and one trained with plain SGD without privacy considerations, especially for complex tasks [12]. Further improvement of DP-SGD turns out to be a highly challenging task: apart from the tighter privacy analysis in [33] mentioned above, to our knowledge, there has not been an effective algorithmic improvement of the core DP-SGD algorithm ever since its proposal [2] in 2016. Instead, later works have focused on other aspects of model building such as noise decay scheduling [64], neural network layers [12, 42], feature engineering [52], utilization of an additional public dataset [39], hyperparameter tuning [41], and efficiency enhancements [50]. These approaches are typically built upon the DP-SGD optimizer; consequently, the unimproved core of DP-SGD have increasingly become a fundamental limiting factor for the performance of DP-compliant deep learning models.

Motivated by this, we propose DPIS, which directly improves upon DP-SGD on the core optimizer level. DPIS can be used as a drop-in replacement of DP-SGD, and can be used in combination with the above-mentioned methods that internally call DP-SGD as the underlying optimizer. The main idea of DPIS is to incorporate *importance sampling (IS)* in each SGD iteration, which replaces the random sampling step in non-private SGD and DP-SGD. We consider the norm of gradient of each record as its *importance* and sample the record with probability proportional to it. It is well-known that in the non-private setting, IS helps reduce sampling variance and accelerate the convergence rate of SGD [11, 21, 66]. More importantly, as explained later in Section 3, the use of IS also helps reduce the amount of noise injected to the gradients that is required to satisfy differential privacy. However, integrating IS into DP-SGD is highly non-trivial as it significantly complicates the (already rather complex) privacy loss analysis. In addition, as explained later in Section 3, IS requires the release of additional statistics which needs to be protected under differential privacy. Finally, a straightforward implementation of IS involves gradient computation for each record in every iteration, which can be computationally infeasible for large training sets.

DPIS addresses the above challenges with novel algorithmic designs accompanied by precise privacy loss analysis. Specifically, in each iteration, DPIS samples each record with probability proportional to its gradient norm, and weights the gradients to obtain an unbiased estimation of the aggregate gradient. To avoid the time-consuming gradient computation for each individual record, we split the IS process into two steps, and apply rejection sampling to reduce the computation cost without significantly compromising result quality. The new private statistics required by IS is carefully perturbed to satisfy DP, and the privacy loss they incur is small. In our privacy analysis, we leverage the results in [33] and prove that the privacy cost of DPIS is strictly lower than that of DP-SGD, which establishes the theoretical basis of the benefits of applying DPIS compared to DP-SGD. As a further optimization, we incorporate into DPIS an adaptive gradient clipping bound update schedule, to further improve the utility of the resulting model.

We have performed a thorough set of experiments, using four classic benchmark datasets: MNIST, Fashion MNIST (FMNIST), CIFAR-10, and IMDb. The former three correspond to computer vision tasks for which we apply convolutional neural networks, whereas the latter one involves natural language processing, for which we use a recurrent neural network. The evaluation results demonstrate that DPIS consistently and significantly outperforms DP-SGD, as well as several recent solutions, under a wide range of hyperparameter settings.

To summarize, the main contributions of this paper include: (i) DPIS, a novel differentially private SGD optimizer that incorporates importance sampling to reduce sampling variance and the injected noise required to satisfy DP, (ii) rigorous privacy loss analysis that proves the superiority of DPIS over DP-SGD, (iii) an adaptive gradient clipping bound updating schedule, and (iv) extensive experiments that evaluate the effectiveness of DPIS on benchmark datasets and common neural network architectures.

# 2 PRELIMINARIES

## 2.1 Differential Privacy

Differential privacy (DP) [14] provides a rigorous, information-theoretic privacy guarantee to prevent an adversary from inferring the presence or absence of any individual record. We go through the basics of DP in the following.

*Definition 2.1 (Neighbor Datasets).* Two datasets $D$ and $D'$ are *neighbor datasets*, if and only if they differ in one single record.

*Definition 2.2 $((\epsilon, \delta)$-Differential Privacy (DP) [16]).* A randomized mechanism $\mathcal{M} : \mathcal{D} \to \mathcal{R}$ with domian $\mathcal{D}$ and range $\mathcal{R}$ satisfies $(\epsilon, \delta)$-DP if

$$\Pr[\mathcal{M}(D) \in \mathcal{O}] \le \exp(\epsilon) \cdot \Pr[\mathcal{M}(D') \in \mathcal{O}] + \delta, \qquad (1)$$

for any set of output $\mathcal{O} \subseteq \mathcal{R}$ and any neighbor datasets $D, D' \subseteq \mathcal{D}$.

Equation (1) enforces an upper bound on the probability ratio of all possible outcomes from two neighbor datasets. Its rationale is that an adversary observing the output $\mathcal{O}$ of $\mathcal{M}$ would have limited confidence to tell if the input is a certain dataset $D$ or its neighbor $D'$. In other words, an adversary is unlikely to infer whether the information of any individual $x$ in the input $D$. The parameter $\epsilon$ is known as the *privacy budget* that affects the adversary's confidence level for telling $D$ from its neighbor $D'$ given the outputs of a mechanism $\mathcal{M}$. The parameter $\delta$ is usually set to a small value representing the probability that the privacy guarantee of $\epsilon$ fails.

---
[1]https://github.com/tensorflow/privacy

An alternative notion of privacy guarantee is Rényi differential privacy (RDP) [32], which is defined based on Rényi divergence [54], as follows.

*Definition 2.3 (Rényi Divergence).* For two probability distributions $\mathcal{P}, \mathcal{Q} \subseteq \mathcal{R}$ and their repective densities $p(x), q(x), x \in \mathcal{D}$, the Rényi divergence of a finite order $\alpha > 1$ is defined as

$$D_\alpha(P\|Q) \triangleq \frac{1}{\alpha - 1} \ln \int_{\mathcal{X}} q(x) \left( \frac{p(x)}{q(x)} \right)^\alpha dx.$$

*Definition 2.4 (($\alpha, \tau$)-Rényi Differential Privacy (RDP)).* A randomized mechanism $\mathcal{M} : \mathcal{D} \rightarrow \mathcal{R}$ with domain $\mathcal{D}$ and range $\mathcal{R}$ satisfies ($\alpha, \tau$)-*RDP* if

$$D_\alpha\left(\mathcal{M}(D)\|\mathcal{M}\left(D'\right)\right) \leq \tau$$

for any neighbor datasets $D, D' \subseteq \mathcal{D}$.

Compared to the ($\epsilon, \delta$)-DP definition (Definition 1), the above RDP notion enables a quantitatively tighter way to track cumulative privacy loss, which leads to more precise privacy analysis for differentially private deep learning, as elaborated later in this section.

Given specific $\delta$, we can convert an ($\alpha, \tau$)-RDP mechanism to one enforcing ($\epsilon, \delta$)-DP, through the following lemma.

LEMMA 2.5 (($\alpha, \tau$)-RDP TO ($\epsilon, \delta$)-DP CONVERSION [7]). *Consider a randomized mechanism $\mathcal{M} : \mathcal{D} \rightarrow \mathcal{R}$ with domain $\mathcal{D}$ and range $\mathcal{R}$. Let $\alpha > 1$ and $\epsilon \geq 0$. If $\mathcal{M}$ satisfies ($\alpha, \tau$)-RDP, given a specific $\delta$, $\mathcal{M}$ satisfies ($\epsilon, \delta$)-DP for*

$$\varepsilon = \tau + \frac{\log(1/\delta) + (\alpha - 1)\log(1 - 1/\alpha) - \log(\alpha)}{\alpha - 1}. \quad (2)$$

Next, we present the concept of *sensitivity*, which is used in basic building blocks of differentially private mechanisms. Let $f$ be a function that maps a dataset from a domain $\mathcal{D}$ into a $d$-dimensional output in $\mathbb{R}^d$. To convert $f$ into a differentially private mechanism, a canonical approach is to inject random noise into the output of $f$, whose scale is calibrated according to the sensitivity of $f$, defined as follows.

*Definition 2.6 (Sensitivity [16]).* The sensitivity $S(f)$ of a function $f : \mathcal{D} \rightarrow \mathbb{R}^d$, denoted as $S(f)$, is defined as

$$S(f) = \max_{D, D'} \|f(D) - f(D')\|,$$

where $D, D'$ denotes two neighbor datasets, and $\|\cdot\|$ is a norm.

In this paper, we focus on the case that $\|\cdot\|$ is the $L_2$ norm; the corresponding sensitivity definition is known as $L_2$ sensitivity.

In DP-SGD and the proposed solution DPIS, a fundamental building block is the *sampled Gaussian mechanism* (SGM) [33], which operates by sampling a subset uniformly at random from a given dataset, applies a function $f$ on the sample set, and injects spherical Gaussian noise according to sensitivity $S(f)$. SGM ensures ($\alpha, \tau$)-RDP according to the following lemma.

LEMMA 2.7 (PRIVACY OF SAMPLED GAUSSIAN MECHANISM [33]). *Given a function $f$ with sensitivity $S(f)$, SGM with sampling probability $p$ and additive Gaussian noise $\mathcal{N}(0, \sigma^2 S(f)^2 \mathbb{I})$ ensures ($\alpha, \tau$)-RDP, where*

$$\tau \leq D_\alpha\left((1 - p)\mathcal{N}\left(0, \sigma^2\right) + p\mathcal{N}\left(1, \sigma^2\right) \| \mathcal{N}\left(0, \sigma^2\right)\right). \quad (3)$$

*When $p = 1$, i.e., every record is included in the sample set and involved in the computation of the output of $f$, SGM reduces to the Gaussian Mechanism (GM) [32].*

Finally, a nice property of differential privacy is that the privacy guarantee is not affected by postprocessing steps, as stated in the following lemma.

LEMMA 2.8 (POSTPROCESSING [17]). *Let $\mathcal{M}$ be a mechanism satisfying ($\epsilon, \delta$)-DP. Let $f$ be a function whose input is the output of $\mathcal{M}$. Then, $f(\mathcal{M})$ also satisfies ($\epsilon, \delta$)-DP.*

The above property also applies to the RDP definition [32].

## 2.2 Sampling Methods

The differential privacy guarantee of a mechanism can be amplified by running it with a randomly sampled subset instead of the whole dataset. In each iteration of differentially private deep learning algorithms, including DP-SGD, they obtain a subset of training samples using Bernoulli sampling.

**Bernoulli sampling.** Bernoulli sampling samples each record by an independent Bernoulli trial with equal probability.

We describe how DP-SGD utilizes Bernoulli sampling works later in Section 2.3.

One idea towards an improved sampling process is to assign an individual sampling probability to each record, rather than having equal sampling probability for all records. Then, the sampling process becomes Poisson sampling. To do so, a main challenge is how to determine the probabilities for the tuples so as to improve the learning efficiency while preserving differential privacy at the same time. Our solution relies on the importance sampling technique, as follows.

**Importance Sampling.** Importance sampling (IS) is a generic sampling technique with applications for Monte Carlo simulations. Specifically, for a Monte Carlo method sampling $g(x), x \in X$ with probability $p(x)$, let $q(x)$ be another probability distribution. The expectation of $g(x)$ can be reformulated as

$$\mathbb{E}_p[g(x)] = \int_X p(x)g(x)dx = \int_X q(x)\frac{p(x)}{q(x)}g(x)dx$$
$$= \mathbb{E}_q\left[\frac{p(x)}{q(x)}g(x)\right],$$

where $\mathbb{E}_p[\cdot]$ and $\mathbb{E}_q[\cdot]$ represent the expectation of $\cdot$ under $p(x)$ and $q(x)$ respectively. The above derivation replaces the underlying probability density function $p(x)$ with $q(x)$ and weights each sample $g(x)$ by $p(x)/q(x)$. It allows us to sample from an easily accessible $q(x)$ to estimate the intentional expectation $\mathbb{E}_p[g(x)]$. We explain how DPIS utilizes IS works later in Section 3.2.

## 2.3 Deep Learning with Differential Privacy

In deep learning, a deep neural network (NN) model is trained with an input training dataset $D$, through an optimizer such as stochastic gradient descent (SGD). Abadi et al. [2] present an $(\epsilon, \delta)$-differentially private algorithm for NN training called DP-SGD. The basic idea is to perturb the gradients in each SGD iteration under differential privacy, and reconstruct the model through these noisy gradients. There are two main challenges for realizing this idea. First, each gradient can be an arbitrary vector with complex dependencies on the input data, leading to unbounded $L_2$ sensitivity (Definition 2.6). Second, the training can involve many iterations, meaning that simply applying a coarse-grained composition [15] or the even the advanced composition [22]) leads to overwhelming noise required to satisfy DP. DP-SGD addresses both issues, explained below.

Similar to non-private SGD, DP-SGD starts by initializing the trainable weights $\theta$ in the NN with random values $\theta = \theta_0$, and iteratively updates $\theta$ with the goal of minimizing a loss function $L(\theta, x)$. In each iteration, DP-SGD draws a batch of records from the training data by Bernoulli sampling. Unlike non-private SGD, DP-SGD additionally ensures the following condition in the sampling: that in each iteration $t$, for a given expected batch size $b$, the batch $\mathcal{B}_t$ is obtained by sampling each tuple $x_i \in D$ independently with a fixed probability $p_i = b/N$, where $N = |D|$. Thus, $\mathcal{B}_t$ has an expected size $\mathbb{E}[|\mathcal{B}_t|] = \sum_{x_i}^D p_i = b$. Then, the method computes the gradient $g_t(x_i)$ for each tuple $x_i \in \mathcal{B}_t$ as $g_t(x_i) = \nabla_{\theta_i} L(\theta_i, x_i)$.

To bound the sensitivity of the gradient, DP-SGD enforces another condition: an upper bound $C$ on the $L_2$ norm of each gradient, resulting in the clipped gradient in Equation (4).

$$\bar{g}_t(x_i) = \textbf{Clip}(g_t(x_i); C), \tag{4}$$
$$\text{where } \textbf{Clip}(x; C) \triangleq x/\max\left(1, \frac{\|x\|_2}{C}\right).$$

After gradient clipping, given two neighbor datasets, their sum of gradients in each iteration differ by at most $C$; hence, the sensitivity of $\sum_{i \in \mathcal{B}_t} g(x_i)$ is bounded. Then, DP-SGD perturbs the mean gradient in a batch using the sampled Gaussian mechanism described in the previous subsection:

$$\tilde{g}_t = \frac{1}{b}\left(\sum_{i \in \mathcal{B}_t} \bar{g}_t(x_i) + \mathcal{N}\left(0, \sigma_G^2 C^2 \mathbf{I}\right)\right), \tag{5}$$

where $\sigma_G$ is the noise multiplier depending on the privacy budget, and $\mathbf{I}$ is the identity matrix. Finally, DP-SGD updates $\theta$ with $\theta_{t+1} = \theta_t - \eta \tilde{g}_t$.

The sampling and clipping conditions enable a fine-grained privacy loss analysis called moments accountant (MA) for ensuring $(\epsilon, \delta)$-DP. Mironov et al. [33] show that DP-SGD satisfies $(\alpha, \tau)$-RDP, and present a numerical procedure for computing the values of $(\alpha, \tau)$, as stated in Lemma 2.9.

**Lemma 2.9 (Privacy Cost of DP-SGD).** *Consider the DP-SGD algorithm with sampling probability $p = b/N$ and noise multiplier $\sigma_G$. Given an integer $\alpha > 1$, each iteration of DP-SGD satisfies $(\alpha, \tau)$-RDP where*

$$\tau = \frac{1}{\alpha - 1}\ln\left(\sum_{m=0}^{\alpha}\binom{\alpha}{m}\left(1 - \frac{b}{N}\right)^{\alpha - m}\left(\frac{b}{N}\right)^m \exp\left(\frac{(m^2 - m)}{2\sigma_G^2}\right)\right). \tag{6}$$

To prove that DP-SGD satisfies $(\epsilon, \delta)$-DP, one can apply the conversion presented in Lemma 2.5. This analysis technique, which first establishes $(\alpha, \tau)$-RDP of the algorithm and then converts it to $(\epsilon, \delta)$-DP, leads to more precise computation of the privacy loss compared to the MA technique in the original DP-SGD paper [2]. Accordingly, it is preferable to compute the noise scale based on the analysis in [33] and Lemma 2.5, which result in lower noise and higher model utility. Hence, the proposed solution DPIS, presented in the next section, follows this analysis technique.

## 3 DPIS

### 3.1 Solution Overview

Recall from Section 2.3 that each iteration of DP-SGD consists of the following operations (i) forming a batch by sampling each tuple uniformly, (ii) computing and clipping the gradient with respect to each sample record, (iii) perturbing each gradient by injecting random Gaussian noise guided by the privacy loss analysis, and (iv) updating the model weights with the averaged noisy gradients in the batch. DP-SGD guarantees that each perturbed gradient is unbiased with respect to the exact value of the clipped gradient defined in Equation (4) [2]. Accordingly, the quality of the model update mainly depends upon the *variance* of the noisy gradients. The proposed solution DPIS follows a similar framework, and aims at improving model utility by reducing the variance of the perturbed gradients in each iteration, while maintaining their unbiasedness.

More precisely, each iteration of DP-SGD aims to update the model with an estimated mean gradient for the entire training set, which is computed from the sampled batch with additional Gaussian noise as explained before. The variance of this estimated mean gradient comes from two sources: sampling noise (which also exists in the non-private setting), and the additional Gaussian noise required to satisfy differential privacy. We focus on reducing the latter through importance sampling (IS) for batch selection, which is a novel contribution of this paper. We take the gradient norm of each record as its importance. Note that IS in general is also known to reduce sampling variance and accelerate the convergence of stochastic optimization in the machine learning literature; we refer the reader to Ref. [21, 23, 66] for a discussion on this topic.

In particular, DPIS samples each tuple with probability proportional to the $L_2$ norm of its gradient. Clearly, this scheme is biased towards records with high gradient norm values. To obtain an unbiased estimate of the mean gradient with respect to the clipped value over the entire training set, we introduce a scaling factor for each gradient, such that all scaled gradients have the same $L_2$ norm. The estimated mean gradient is then computed as the average of the scaled gradients over all records in batch sampled through IS. Later in Section 4, we provide rigorous analysis on the privacy loss of DPIS, and prove that it satisfies both RDP and $(\epsilon, \delta)$-DP.

In the following, Section 3.2 provides the theoretical rationale of integrating IS into the DP-SGD framework, in which we formulate the choice of the sampling strategy as an optimization program.

The use of IS introduces two complications to the algorithm design: (i) IS requires the release of private statistical information that must be protected under DP, and (ii) IS involves the computation of the gradient for every record in the training data, which can be infeasible for a large training set. Section 3.3 deals with (i), and analyzes the privacy cost of the additional data release. Section 3.4 tackles (ii) through a novel rejection sampling process, which significantly reduces the computational cost of DPIS.

In addition, we provide an optimization for DPIS in Section 3.5, which applies a novel adaptive update schedule for the gradient clipping upper bound $C$ (Equation (4)) during training, instead of fixing $C$ to a constant as is done in DP-SGD. Finally, Section 3.6 summarizes the complete DPIS algorithm.

## 3.2 Rationale of Applying IS

Recall from Section 2.3 that given an expected batch size $b$, DP-SGD samples each record $x_i$ with a fixed probability $b \cdot \frac{1}{N}$, where $N$ is the size of the training set $D$. In this subsection, we demonstrate through theoretical analysis that this sampling scheme is suboptimal in terms of noise variance, and derive an improved sampling scheme with IS.

Without loss of generality, consider a single iteration during SGD training. Let $b \cdot p(x_i)$ be the probability for sampling $x_i$, in which $p$ is called the *sampling probability distribution*. Clearly, in DP-SGD, we have $p(x_i) = \frac{1}{N}$ for all $x_i$. Let $\mathcal{B}$ denote the batch obtained through sampling. We enforce the condition $\sum_i^N p(x_i) = 1$, which ensures that the expected size of $\mathcal{B}$ is $\mathbb{E}[|\mathcal{B}|] = \sum_i^N b \cdot p(x_i) = b$. In a nutshell, our goal to determine the best values for each $p(x_i)$ such that the estimated mean gradient remains unbiased, and at the same time the variance of the estimated mean gradient is minimized. This is done through solving an optimization program, described below.

**Maintaining unbiasedness.** For gradient descent methods [45], which are commonly used in machine learning, using an unbiased estimate of the gradient mean leads to guranteed convergence of parameters, when the number of iterations approaches infinity [4, 55]. In the non-private setting, popular optimizers ensure the unbiasedness of the gradient mean, e.g., through explicit bias-correction in Adam. DPIS follows this common practice. To obtain an unbiased estimation of the mean gradient with respect to the clipped value in the training dataset, we scale each gradient $g(x_i)$ with the factor $\frac{1}{Nbp(x_i)}$, and output the sum of the scaled gradients to estimate the mean gradient. We have:

$$\mathbb{E}[g] = \frac{1}{N}\sum_i^N g(x_i) = \sum_i^N bp(x_i) \cdot \frac{1}{Nbp(x_i)}g(x_i).$$

Accordingly, the estimated mean clipped gradient using the sampled batch $\mathcal{B}$ (corresponding to Equation (5) in DP-SGD) becomes:

$$\tilde{g} = \frac{1}{b}\left(\sum_{i\in\mathcal{B}}\frac{1}{Np(x_i)}\bar{g}(x_i) + \mathcal{N}\left(0, \sigma_G^2 C^2 \mathbf{I}\right)\right). \tag{7}$$

where $\bar{g}$ is the clipped gradient defined in Equation (4), and $\tilde{g}$ is an unbiased estimator for the mean value of $\bar{g}$ over the entire training set, according to the following equation.

$$\mathbb{E}[\tilde{g}] = \mathbb{E}\left[\sum_{i\in\mathcal{B}}\frac{1}{Nbp(x_i)}\bar{g}(x_i)\right] + \frac{1}{b}\mathbb{E}\left[\mathcal{N}(0, \sigma_G^2 C^2 \mathbf{I})\right] = \frac{1}{N}\sum_i^N \bar{g}(x_i).$$

In the following, to simplify our notations, we use symbol $Z$ to denote the injected Gaussian noise required to satisfy differential privacy, i.e., $Z$ follows a Gaussian distribution of $\mathcal{N}(0, \sigma_G^2 C^2 \mathbf{I})$. Accordingly, Equation (7) simplifies to $\tilde{g} = \frac{1}{b}\left(\sum_{i\in\mathcal{B}}\frac{1}{Np(x_i)}\bar{g}(x_i) + Z\right)$.

**Minimizing noise variance.** Recall from Section 2.3 that the goal of SGD is to compute $\theta^*$ that minimizes the loss function $L(\theta^*)$, i.e., $\theta^* \triangleq \arg\min_\theta L(\theta)$. According to Theorem 2 of [46], the accuracy of this empirical risk minimization task depends primarily on the expected squared $L_2$ norm of each gradient update. Here, we aim to minimize $\mathbb{E}[\|\tilde{g}\|^2]$ by calibrating the sampling probability distribution $p$.

We denote $\mathcal{B}$ as a batch of tuples and $p(\mathcal{B})$ as the corresponding probability that $\mathcal{B}$ is sampled. By definition, we have

$$p(\mathcal{B}) = \prod_{i\in\mathcal{B}} bp(x_i) \prod_{l\in D\setminus\mathcal{B}}(1 - bp(x_l)). \tag{8}$$

We let $d$ be the number of parameters in the machine learning model and let $Z$ follow a Gaussian distribution of $\mathcal{N}(0, \sigma_G^2 C^2 \mathbf{I}_d)$. Note that $\mathbb{E}[\|Z\|^2] = \sigma_G^2 C^2 d$. Then, our objective $\mathbb{E}[\|\tilde{g}\|^2]$ can be expressed as

$$\mathbb{E}[\|\tilde{g}\|^2] = \frac{1}{b^2}\sum_\mathcal{B} p(\mathcal{B})\mathbb{E}_Z\left[\|\sum_{i\in\mathcal{B}}\bar{g}(x_i) + Z\|^2\right]$$

$$= \frac{1}{b^2}\mathbb{E}[\|Z\|^2] + \frac{1}{b^2}\sum_\mathcal{B}\prod_{i\in\mathcal{B}}bp(x_i)\prod_{l\in D\setminus\mathcal{B}}(1 - bp(x_l))\cdot$$

$$\left(\sum_{i\in\mathcal{B}}\frac{\|\bar{g}(x_i)\|^2}{N^2 p(x_i)^2} + \sum_{i\neq j\in\mathcal{B}}\frac{\bar{g}(x_i)^T\bar{g}(x_j)}{N^2 p(x_i)p(x_j)}\right)$$

$$= \frac{1}{b^2}\mathbb{E}[\|Z\|^2] + \frac{1}{bN^2}\sum_i^N\frac{\|\bar{g}(x_i)\|^2}{p(x_i)} + \frac{1}{N^2}\sum_{i\neq j}^N\bar{g}(x_i)^T\bar{g}(x_j),$$

The first equality expands $\mathbb{E}[b^2\|\tilde{g}\|^2]$ to all possible outcomes w.r.t batch selections and $\mathbb{E}_Z$ is the expectation w.r.t the variable $Z$. The second equality is due to the fact that $Z$ is unbiased and independent with the outcome of $p(x_i)$. The last equality is that: For $\forall i\in D$, the appearance of $\frac{\|\bar{g}(x_i)\|^2}{N^2 p(x_i)^2}$ only depends on whether $i$ is sampled into $\mathcal{B}$, no matter whether the others are sampled. Hence, the sum of probability products of each batch containing $i$ is $bp(x_i)$; Similarly, the sum of probability products of each batch containing $i, j$ is $b^2 p(x_i)p(x_j)$.

Observe that $\sum_{i\neq j}^N \bar{g}(x_i)^T\bar{g}(x_j)$ is a constant w.r.t. $\{p(x_i), p(x_j)\}$. Thus, to minimize $\mathbb{E}[\|\tilde{g}\|^2]$, it suffices to derive

$$\arg\min_p \frac{1}{bN^2}\sum_i^N\frac{\|\bar{g}(x_i)\|^2}{p(x_i)} + \frac{1}{b^2}\sigma_G^2 C^2 d. \tag{9}$$

**Choosing $p$.** In Equation (9), $\sigma_G$ in the second term has a rather complicated dependency on $p$, which is difficult to optimize. We defer the analysis of $\sigma_G$ to Section 4. Here, we focus on minimizing the first term, which represents the sample variance. Since $f(x) = x^2$ is convex, we apply Jensen's inequality and obtain

$$\sum_i^N \frac{\|\bar{g}(x_i)\|^2}{p(x_i)} = \sum_i^N p(x_i) \left( \frac{\|\bar{g}(x_i)\|}{p(x_i)} \right)^2 \geq \left( \sum_i^N p(x_i) \frac{\|\bar{g}(x_i)\|}{p(x_i)} \right)^2,$$

where the equality holds when all $\|g(x_i)\|/p(x_i)$ are equal. Hence, an appropriate choice of $p$ that minimizes the sample variance is

$$p(x_i) = \frac{\|\bar{g}(x_i)\|}{K}, \text{ where } K = \sum_i^N \|\bar{g}(x_j)\|.$$

The above choice of $p$ corresponds to the IS scheme used in DPIS, *i.e.*, sampling each tuple $x_i$ proportionally to the $L_2$ norm of its gradient. Our analysis so far focuses on minimizing sample variance, *i.e.*, the first term of Equation (9). It turns out that the IS strategy also helps reduce the variance of the injected Gaussian noise (the second term of Equation (9)), explained in the next subsection.

## 3.3 Private IS

In each SGD iteration, apart from the estimated mean gradient, DPIS needs to release two additional values computed from private data, due to the use of importance sampling: (i) the total number $N$ of records in the training set, and (ii) the sum of gradient norms $K = \sum_i^N \|\bar{g}(x_i)\|$. To satisfy differential privacy, DPIS perturbs $N$ and $K$ with Gaussian noise as follows:

$$\tilde{N} = N + \mathcal{N}\left(0, \sigma_N^2\right), \tag{10}$$

$$K' = \frac{1}{p_K} \sum_{i \in \mathcal{B}_K} \|\bar{g}(x_i)\| + \mathcal{N}(0, \sigma_K^2 C^2), \tag{11}$$

where $\mathcal{B}_K$ denotes a batch, and parameters $\sigma_N$ and $\sigma_K$ control the variance of the Gaussian noise. The noisy values $\tilde{N}$ and $K'$ are then released alongside the noisy gradient sum $\tilde{g}$ defined in Equation (7), in each iteration of DPIS. In the above definitions of $\tilde{N}$ and $K'$, Equation (10) is straightforward as it directly follows the Gaussian mechanism; meanwhile, in Equation (11), we leverage subsampling [3] to reduce the privacy cost of releasing $K'$: We uniformly sample a batch $\mathcal{B}_K$ of records with probability $p_K$ and use the batch to estimate the gradient sum of all records.

Given $K'$, DPIS refines it by imposing an upper bound $\tilde{N} \cdot C$ and a lower bound $b \cdot C$ as follows:

$$\tilde{K} = \min\left(\max\left(K', bC + \xi\right), \tilde{N}C\right), \tag{12}$$

where $\xi$ is a small positive constant. To explain, recall that each gradient has a clipped $L_2$ norm that is at most $C$, according to Equation (4). Therefore, the exact value of $K$ is no larger than $NC$ and, thus, $\tilde{N}C$ is a reasonable upper bound for $\tilde{K}$. Meanwhile, as we discuss in Section 3.2, the sampling probability of each record $x_i$ is $b \cdot p(x_i) = \frac{b\|\bar{g}(x_i)\|}{\tilde{K}} \leq \frac{bC}{\tilde{K}}$. This probability should be less than 1; otherwise, there is no privacy amplification by sampling, resulting in a high privacy cost. Hence, $\tilde{K}$ cannot be arbitrarily small. Motivated by this, we impose a lower bound on $\tilde{K}$ in Equation (12). The definitions of $\tilde{N}$ and $\tilde{K}$ also involve two parameters: $\sigma_N$ and $\sigma_K$. Their values depend on the privacy cost analysis of releasing $\tilde{g}$, $\tilde{K}$, and $\tilde{N}$, which we defer to Section 4 for ease of exposition.

Recall that from Section 3.2 that, when applying IS, we choose to sample each record with probability proportional to its gradient norm for the purpose of minimizing the sample variance, *i.e.*, the first term of Equation (9). As shown in the following lemma, this IS scheme also reduces the privacy cost of releasing gradients, which, in turn, reduces the amount of noise required in the gradients to satisfy differential privacy.

**Lemma 3.1 (Privacy cost of DPIS).** *Consider an iteration of the DPIS algorithm with noisy values $\tilde{N}$, $\tilde{K}$, sampling probability $\frac{b\|\bar{g}(x_i)\|}{\tilde{K}}$ for each record $x_i$, clipping threshold $C$, and noise multiplier $\sigma_G$. Given any integer $\alpha > 1$, this DPIS iteration satisfies $(\alpha, \tau)$-RDP where*

$$\tau_{DPIS} = \frac{1}{\alpha - 1} \cdot$$
$$\ln\left( \sum_{m=0}^{\alpha} \binom{\alpha}{m} \left(1 - \frac{bC}{\tilde{K}}\right)^{\alpha-m} \left(\frac{bC}{\tilde{K}}\right)^m \exp\left( \frac{(m^2 - m)\tilde{K}^2}{2\tilde{N}^2 \sigma_G^2 C^2} \right) \right). \tag{13}$$

The proof of the above lemma is deferred to Section 4, which presents the privacy analysis of DPIS. Observe that the above defining equation of $\tau_{DPIS}$ is similar to that of $\tau_{DP-SGD}$ (*i.e.*, Equation (6)), except that $\tilde{N}$ (in DP-SGD) is replaced by $\tilde{K}/C$ (in DPIS). In particular, $\tau_{DPIS}$ increases with $\tilde{K}$, and it becomes identical to $\tau_{DP-SGD}$ when $\tilde{K} = \tilde{N}C$ (assuming that other parameters are identical for the two methods). Note that $\tilde{N}C$ is the upper bound of $\tilde{K}$ according to Equation (12). Therefore, given the same privacy budget $\tau_{DPIS} = \tau_{DP-SGD}$, the Gaussian noise scale $\sigma_G$ used in DPIS is no larger than that used in DP-SGD. The key of why smaller $\tilde{K}$ in DPIS translates to lower privacy loss is that, unlike previous methods that simply sum up the sampled gradients as in Equation (5), DPIS assigns a *weight* to each sampled gradient that is inversely proportional to its sampling probability, as shown in Equation (7). Therefore, when $\tilde{K}$ is large, although a record $x$ with a larger gradient norm has a higher probability to be sampled (which increases its privacy cost), the weight for $x$ also becomes smaller according to Equation (7), which leads to a lower sensitivity that more than compensates for the increase of privacy loss due to $x$'s higher sampling probability.

In practice, the gradient norm values usually decrease as the training processes and gradually converges. As a result, the value of $\tilde{K}$ (and hence, $\tilde{K}/C$) decreases over time, leading to decreasing values of the noise scale $\sigma_G$ in DPIS given a constant privacy budget $\tau_{DPIS}$. In contrast, in DP-SGD, the noise depends on $\tilde{N}$, which remains constant throughout the training iterations. To illustrate this, Figure 1 plots the value of $\sigma_G$ as a function of training epochs, for DPIS and three competitors on the FMNIST dataset with $\epsilon = 0.5, 1, 4, \delta = 10^{-5}$. (Detailed experimental settings are described later in Section 5.) Observe that, when $\epsilon = 4$, the noise multiplier $\sigma_G$ of DPIS gradually decreases from 0.81 to 0.42, whereas DP-SGD requires a constant noise multiplier $\sigma_G = 0.81$. The decrease in gradient noise at later epochs is particularly helpful, since the gradient values tend to become smaller and more sensitive to noise as the training approaches convergence.

Summarizing the analysis in Sections 3.2 and 3.3, the novel sampling scheme in DPIS reduces both sample variance (as explained in Section 3.2) and the scale of the injected noise (as described in this subsection), which optimizes both terms in the objective defined in Equation (9). Hence, DPIS is a direct improvement over the classic

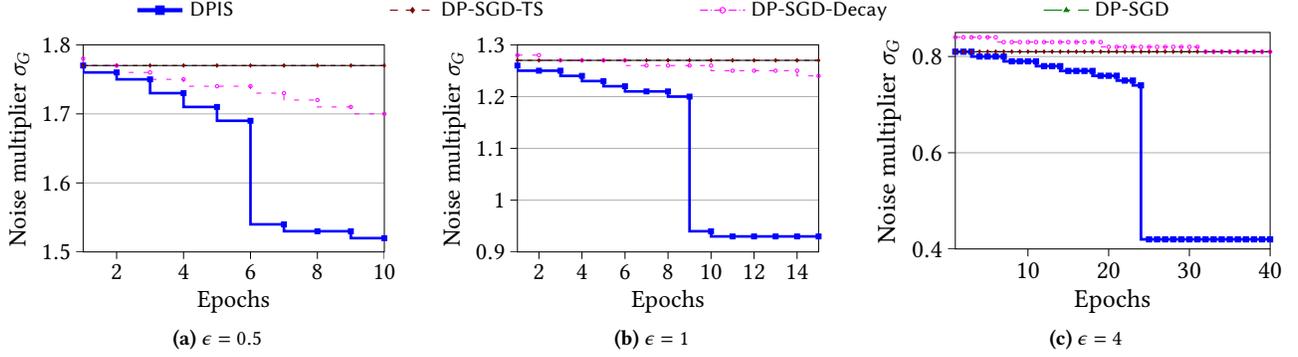

**Figure 1: Noise multipliers of DPIS and three competitors on FMNIST, with total privacy budge $\epsilon \in \{0.5, 1, 4\}$ and $\delta = 10^{-5}$.**

DP-SGD optimizer as the former incurs lower noise in the released gradients.

### 3.4 Accelerating Private IS with Rejection Sampling

So far in our descriptions, the sampling scheme in DPIS requires the knowledge of the gradient value for every record in the dataset, which increases the computational overhead for each iteration from $O(b)$ to $O(N)$, where $b$ and $N$ are the sample size and dataset cardinality, respectively. This is rather inefficient, and can become computationally infeasible for a massive training set with a large $N$. To deal with this problem, DPIS introduces an additional filtering step (before applying IS) based on rejection sampling, detailed below.

**Rejection Sampling.** Suppose that we have an unknown target distribution function $p(x) = \tilde{p}(x)/K_p$, $\sum_x^D p(x) = 1$, where $\tilde{p}(x)$ is known and the $K_p$ is an unknown normalization constant. The idea of rejection sampling is that when we cannot sample from the $p(x)$ directly, we can instead use a proxy distribution function $q(x)$, $\sum_x^D q(x) = 1$. The method introduces a constant $k$ such that $kq(x) \geq \tilde{p}(x)$ is satisfied for all $x \in D$. The steps for rejection sampling are as follows:

(1) The algorithm independently samples each $x \in D$ with probability $q(x)$.

(2) Let $\mathcal{X}_q$ the set of tuples sampled in step (1). For each $x \in \mathcal{X}_q$, the method uniformly samples a $u$ in $[0, kq(x)]$. If $u \leq \tilde{p}(x)$, accept the tuple $x$; otherwise, reject $x$.

The algorithm essentially accepts each $x$ with probability

$$q(x) \cdot \frac{\tilde{p}(x)}{kq(x)} = \frac{\tilde{p}(x)}{k}. \tag{14}$$

In general, to avoid rejecting numerous samples, which results in low sampling efficiency, the constant $k$ should be as small as possible, subject to $kq(x) \geq \tilde{p}(x)$.

Specifically, DPIS first selects a subset from the dataset following a probability distribution $q$ (which is efficient to compute), and then performs IS with the actual gradients (which are expensive to compute) on the subset instead of the whole dataset. To realize this, there are two issues that need to be addressed. The first issue due to the additional filtering step performed with rejection sampling, the batch collected with IS may contain fewer samples than the expected batch size $b$. In particular, recall from Equation (14) that the expected number of acceptance in rejection sampling is

$$\mathbf{E}[acceptance] = \sum_{x \in D} \frac{\tilde{p}(x)}{k} = \frac{K_p}{k} \sum_{x \in D} p(x) = \frac{K_p}{k},$$

where $K_p = \sum_{x \in D} \tilde{p}(x) \leq \sum_{x \in D} kq(x) = k$. The above equation indicates that after rejection sampling, our actual sampling probability for each tuple is scaled by a factor of $K_p/k \leq 1$. Consequently, the resulting batch contains fewer than $b$ samples, leading to higher sampling variance. To remedy this, we modify the filtering step to sample each tuple with probability $\min(kq(x), 1)$ instead of $q(x)$.

Second, according to the IS scheme, we generally prefer data records with larger gradient norms; hence, we need to construct an effective and efficient filter $q$ for gathering such records. To do so, DPIS employs the following strategy: at the beginning of each epoch (which contains multiple iterations), we compute the gradient $g_0(x)$ for each tuple $x$ as a factor of the numerator of $q(x)$, and then compute the noisy gradient sum $\tilde{K}$ as the denominator of $q(x)$; when we sample $x$ in a certain iteration, we update the numerator of $q(x)$ with the latest gradient $g_t(x)$.

Let $T = N/b$ be the number of iterations in one training epoch. The complete sampling process of DPIS for iteration $t \in [1, T]$ in any epoch is as follows:

(1) If $t = 1$ (*i.e.*, the first iteration in an epoch), we compute gradient $g_0(x_i)$ and clip with the gradient norm bound $C$ by Equation (4) for each $x_i \in D$. Then, we release a noisy gradient sum $\tilde{K}$ by Equation (12). Given a pre-set constant $k$ and a lower bound $g_L \ll C$, we compute the following value:

$$\hat{g}_1(x_i) = k \max(\|\tilde{g}_0(x_i)\|, g_L).$$

(2) For each $t \in [1, T]$, let

$$q_t(x_i) = \min\left(\frac{b \cdot \hat{g}_t(x_i)}{\tilde{K}}, 1\right).$$

We sample tuple $x_i$ independently with probability $q_t(x_i)$. Let $\mathcal{X}_q$ be the set of tuples sampled by $q_t$.

(3) For $x_i \in \mathcal{X}_q$, we compute its gradient $g_t(x_i)$, clip with upper bound $\min(\hat{g}_t(x_i), C)$ by Equation (4) and accept it with a second round probality

$$p_t(x_i) = \frac{\|\tilde{g}_t(x_i)\|}{\hat{g}_t(x_i)}.$$

Let $\mathcal{X}_p$ be the set of accepted tuples. We use $\mathcal{X}_p$ to perform gradient descent as illustrated in Section 3.6.

(4) Lastly, we update the $\hat{g}_{t+1}(x_i)$ for all $x_i \in \mathcal{X}_q$, i.e.,

$$\hat{g}_{t+1}(x_i) = k \max(\|\bar{g}_t(x_i)\|, g_L).$$

In Step (2), the lower bound $g_L$ is set to give a chance to sample the tuples whose gradients are too small in previous iterations. For example, if the gradient of one tuple in some iteration is close to **0**, we will never sample it as $q_t(x_i) \approx 0$ without a lower bound. In practice, we usually have $\tilde{K} > kbC$ as long as $N \gg kb$. Then the actual sampling probability of $x_i$ is displayed in Equation (15), which is consistent with our IS scheme.

$$q_t(x_i) \cdot p_t(x_i) = \frac{b\|\bar{g}_t(x_i)\|}{\tilde{K}}. \tag{15}$$

Assuming that $g_L$ is small and the sum of gradients does not vary too much in one epoch, we approximate the expected sizes of $\mathcal{X}_q, \mathcal{X}_p$ using $kb$ and $b$, respectively, based on the following derivations:

$$\mathbb{E}[|\mathcal{X}_q|] = \sum_i^N q_t(x_i) = kb \frac{\sum_i^N \max(\|\bar{g}_{t-1}(x_i)\|, g_L)}{\tilde{K}} \approx kb,$$

$$\mathbb{E}[|\mathcal{X}_p|] = b \sum_i^N \frac{\|\bar{g}_t(x_i)\|}{\tilde{K}} \approx b.$$

Thus, we only compute the gradients of tuples in $\mathcal{X}_q$ and reduce the computation cost from $O(N)$ to $O(kb)$. In addition, the actual batch size used to update the model also approximates the expected size $b$. Overall, we manage to simulate the IS scheme that samples each tuple with probability proportional to its current gradient, at a substantially reduced computational cost.

### 3.5 Adaptive Clipping Bound Updates

DP-SGD uses a constant gradient norm bound $C$ during the whole training process, which is inflexible and has several disadvantages. For instance, if $C$ is (or becomes) significantly larger than the $L_2$ norm of the gradients for most records, the amount of Gaussian noise injected to the gradients (Equation (5)) might be excessive, since the amplitude of the noise is proportional to $C^2$. Moreover, the total training iterations under the DP setting is limited because every additional training step incurs extra privacy costs. Conversely, if $C$ is too small, we would clip gradients aggressively, leading to insufficient updates on model parameters and low training effectiveness. The appropriate value of $C$ may change as training progresses. For example, in our experiments of training an RNN model on IMDb, the mean of gradients increases several times during training. The validation accuracy of the RNN model grows rather slowly under a fixed small $C$, while under a fixed large $C$ the accuracy becomes even worse with more iterations due to overwhelming noise injected to the gradients.

As $C$ increases from 0, the gradient information and the amplitude of the noise grow at the same time. When $C$ increases to a certain extent, the gradient norms of more and more samples are lower than $C$. Inspired by the sum of gradients $\tilde{K}$ in IS, we can approximate the optimal clipping bound by the "mean" of the gradients. However, using gradients or other sensitive information to adjust the clipping bound may consume extra privacy budget.

Hence, we modify $C$ only at the beginning of each epoch so that the privacy cost is negligible. The process of our adaptive clipping is as follows.

(1) In epoch $e \geq 2$, we clip the gradients of tuples in $\mathcal{X}_p$ by an external clipping bound $C^*$ and obtain $K_e^*$ by Equation (12).

(2) We release a noisy version of $K_e^*$ as

$$\tilde{K}_e^* = K_e^* + \mathcal{N}(0, \sigma_K^2 C^{*2}). \tag{16}$$

For epoch $e + 1$, the clipping bound is

$$C_{e+1} = \lambda \cdot \frac{\tilde{K}_e^*}{\tilde{N}}, \tag{17}$$

where $\lambda$ is a hyperparameter served as a quantile.

$\tilde{K}_e^*$ and $C_{e+1}$ change little when $C^*$ varies, because only a small fraction of gradients (whose norms are much larger than others) are clipped by $C^*$. Accordingly, we set $C^* = 4 * C_1$ for all related experiments.

### 3.6 Complete Algorithm

Algorithm 1 outlines pseudocode of the DPIS algorithm under $(\epsilon, \delta)$-DP. For simplicity, we assume that the gradients of all weights and biases in the neural network are stored and processed as one single vector, on which we perform gradient clipping and noise injection. The epoch divider $a_E$ in **Input** is explained in Section 4.2.

Our algorithm starts by initializing the NN model parameters $\theta$ to random values (Line 1). Then, it examines the number $N$ of tuples in $D$, and obtain a noisy version $\tilde{N}$ by injecting Gaussian noise with standard variance $\sigma_N$ (Line 3).

The subsequent execution of the algorithm consists of $E$ epochs, each of which contains $T$ iterations (Lines 4 and 6). Before the iterations, the algorithm initializes several variables including the noisy gradient sum $K_e^*$, and sets $\mathcal{X}_q, \mathcal{X}_p$ (Line 5). At the beginning of each epoch, the algorithm computes gradients for all tuples with clipping applied, computes noisy gradient sum $K_{e,1}$, and generates a list of gradient sum $K_{e,2}, ..., K_{e,T}$ equal to $K_{e,1}$ (Lines 7-14). After obtaining them, the algorithm analyzes the required privacy budget in this epoch and outputs a suitable standard variance $\sigma_{Ge}$ (Line 15) for the additive Gaussian noise in the gradient averaging step (Line 26). Details of **Privacy Analysis** (Lines 2 and 15) is elaborated in Section 4, where we comprehensively discuss how to set noise multipliers $\sigma_N, \sigma_K, \{\sigma_{Ge}\}$ and the use of $a_E$.

In each iteration, the algorithm first samples $x_i \in D$ into $\mathcal{X}_q$ with probability $q_t(x_i)$ (Lines 16-18). For every $x_i \in \mathcal{X}_q$, the algorithm computes its latest gradient with clipping applied, samples it into $\mathcal{X}_p$ with probability $p_t(x_i)$, and updates the numerator $\hat{g}_{t+1}(x_i)$ of $q_{t+1}(x_i)$ (Lines 19-24). (The derivation of $q_t, p_t, \hat{g}_{e,t}, \hat{g}_{e,t+1}$ is discussed in Section 3.4.) Then, the algorithm averages the gradients, injects Gaussian noise, and updates the model parameters (Lines 26 and 27). At the end of each epoch, the algorithm adaptively tunes the gradient norm clipping bound by summing up the recent gradient norms (Line 28) and output a suitable bound $C_{e+1}$ (Line 29).

**Limitations**. In differentially private gradient descent algorithms, computing and storing gradients dominate the running time and memory overheads (without considering the space occupied by the input data), respectively. In each iteration, to obtain a batch of size

---

**Algorithm 1:** Model training with DPIS
**Input:** Dataset $D$ with size $N$, privacy parameters $\epsilon, \delta$, initial gradient clipping bound $C_1$, external bound $C^*$, expected batch size $b$, number of epochs $E$, epoch divider $a_E$, number of iterations $T$, learning rate $\eta$, probability multiplier $k$, gradient lower bound $g_L$, quantile $\lambda$, noise multipliers $\sigma_N, \sigma_K$

**Output:** Model parameter $\theta$.
1  Initialize the model parameter $\theta$ with random values;
2  $\sigma_N, \sigma_K \leftarrow$ **Privacy Analysis**;
3  $\tilde{N} \leftarrow |D| + \mathcal{N}(0, \sigma_N^2)$;
4  **for** $epoch\ e \in [1, E]$ **do**
5     $\tilde{K}_e^* \leftarrow 0;\ \mathcal{X}_q, \mathcal{X}_p \leftarrow \emptyset$;
6     **for** $iteration\ t \in [1, T]$ **do**
7        **if** $t=1$ **then**
8           **for** $x_i \in D$ **do**
9              $g_{e,0}(x_i) \leftarrow \nabla_\theta \mathcal{L}(\theta, x_i)$;
10             $\bar{g}_{e,0}(x_i) \leftarrow \text{Clip}(g_{e,0}(x_i); C_e)$;
11             $\hat{g}_{e,1}(x_i) \leftarrow k \max(\|\bar{g}_{e,0}(x_i)\|, g_L)$;
12             $\bar{g}_e^*(x_i) \leftarrow \text{Clip}(g_{e,1}(x_i); C^*)$;
13          $\tilde{K} \leftarrow$ Equation (12);
14          Generate $\{\tilde{K}_{e,2}, ..., \tilde{K}_{e,T}\}$;
15          $\sigma_{G_e} \leftarrow$ **Privacy Analysis**;
16       **for** $x_i \in D$ **do**
17          $q_{e,t}(x_i) \leftarrow \min(\frac{b \cdot \hat{g}_{e,t}(x_i)}{\tilde{K}_{e,t}}, 1)$;
18          $\mathcal{X}_q \leftarrow \mathcal{X}_q \cup \{x_i\}$ with probability $q_{e,t}(x_i)$;
19       **for** $x_i \in \mathcal{X}_q$ **do**
20          $g_{e,t}(x_i) \leftarrow \nabla_\theta \mathcal{L}(\theta, x_i)$;
21          $\bar{g}_{e,t}(x_i) \leftarrow \text{Clip}(g_{e,t}(x_i); \min(\|\hat{g}_{e,t}(x_i)\|, C_e))$;
22          $p_{e,t}(x_i) \leftarrow \frac{\|\bar{g}_{e,t}(x_i)\|}{\hat{g}_{e,t}(x_i)}$;
23          $\mathcal{X}_p \leftarrow \mathcal{X}_p \cup \{x_i\}$ with probability $p_{e,t}(x_i)$;
24          $\hat{g}_{e,t+1}(x_i) \leftarrow k \max(\|\bar{g}_{e,t}(x_i)\|, g_L)$;
25          $\bar{g}_e^*(x_i) \leftarrow \text{Clip}(g_{e,t}(x_i); C^*)$;
26       $\tilde{g}_{e,t} \leftarrow \frac{1}{b}\left(\sum_{x_i \in \mathcal{X}_p} \frac{1}{N q_{e,t}(x_i) p_t(x_i)} \mathbf{g}_i + \mathcal{N}\left(0, \sigma_{G_e}^2 C_e^2 \mathbf{I}\right)\right)$;
27       $\theta \leftarrow \theta - \eta \tilde{g}_t$;
28    $\tilde{K}_e^* \leftarrow \sum_i^N \|\bar{g}_e^*(x_i)\| + \mathcal{N}(0, \sigma_K^2 C^{*2})$;
29    $C_{e+1} \leftarrow \lambda \cdot \frac{\tilde{K}_e^*}{N}$;
30 **return** $\theta$

---

$b$, DPIS pre-samples $k \cdot b$ records and computes their gradients, where $k$ is a system parameter described in Section 3.4. Therefore, the total computational/memory overhead of DPIS is $k$ times larger than that of DP-SGD. Note that gradient computations for all $k \cdot b$ samples can often be done in parallel on a GPU. In our experiments, we set $k \leq 5$.

# 4 THEORETICAL ANALYSIS

## 4.1 Privacy and Utility Analysis

We first prove that Algorithm 1 ensures $(\epsilon, \delta)$-DP, regarding it as a function with the following input and output:

- Main inputs: $D, \epsilon_0, \delta_0, C_1, C^*, b, T, E, a_E, \sigma_N, \sigma_K$;
- Irrelevant inputs (no impact on privacy analysis): $k, g_L, \eta, \lambda$;

- Outputs: the initial model parameters $\theta$ (Line 1), the noisy version $\tilde{N}$ of $N = |D|$ (Line 3), the first noisy gradient sum $\{\tilde{K}_{e,1}\}, \{\tilde{K}_e^*\}$ in each epoch (Lines 13 and 28), the noisy gradient $\tilde{g}_{e,t}$ in each iteration (Line 26), and the constants $\{\sigma_{G_e}\}$.

We fix $\sigma_N$ and $\sigma_K$ before running the algorithm. We will show that $\theta, \{\sigma_{G_e}\}$ do not incur any privacy cost, and will compute the privacy cost of releasing $\tilde{N}, \{\tilde{K}_{e,1}\}, \{\tilde{K}_e^*\}, \{\tilde{g}_{e,t}\}$ under $(\alpha, \tau)$-RDP and sum them up by the composition theorem of RDP [32]. Based on this sum, we derive the $(\epsilon, \delta)$-DP guarantee Algorithm 1 using Lemma 2.5.

**Privacy analysis for $\theta, \{\sigma_{G_e}\}$.** Note that the initial model parameters $\theta$ are generated randomly, and thus, do not carry any private information. Meanwhile, $\{\sigma_{G_e}\}$ are decided based on the public parameters $\epsilon_0, \delta_0, C_1, C^*, b, E, a_E, T$ and the released parameters $\tilde{N}, \{\tilde{K}_{e,t}\}$, which indicates that they do not reveal any private information in addition to what $\tilde{N}, \{\tilde{K}_{e,t}\}$ divulges. Therefore, the generation of $\theta, \{\sigma_{G_e}\}$ entails zero privacy cost.

**Privacy analysis for $\tilde{N}$.** $\tilde{N}$ is generated by injecting Gaussian noise $\mathcal{N}(0, \sigma_N^2)$ into $N = |D|$ whose sensitivity equals 1. According to Lemma 2.7 with $p = 1$, the privacy cost (in the form of $(\alpha, \tau)$-RDP) of releasing $\tilde{N}$ perturbed by $\mathcal{N}(0, \sigma_N^2)$ is as follows:

$$\tau(\tilde{N}) \triangleq D_\alpha\left(\mathcal{N}(1, \sigma_N^2) \,\|\, \mathcal{N}(0, \sigma_N^2)\right). \tag{18}$$

**Privacy analysis for $\{\tilde{K}_{e,1}\}, \{\tilde{K}_e^*\}$.** The generation of $\{\tilde{K}_{e,1}\}$ and $\{\tilde{K}_e^*\}$ is similar to that of $\tilde{N}$, except a privacy amplification by subsampling and the corresponding sensitivities are changed to $C_e$ and $C^*$, respectively. We fix $p_K = \frac{b}{N}$. By Lemma 2.7 with $p = \frac{b}{N}$, the privacy costs (in the form of $(\alpha, \tau)$-RDP) of releasing $\tilde{K}_{e,1}$ and $\tilde{K}_e^*$ perturbed by $\mathcal{N}(0, \sigma_K^2 C_e^2)$ and $\mathcal{N}(0, \sigma_K^2 C^{*2})$, respectively, are as follows:

$$\tau(\{\tilde{K}_{e,1}\}) \triangleq D_\alpha\left((1-\tfrac{b}{N})\mathcal{N}(0, \sigma_K^2) + \tfrac{b}{N}\mathcal{N}(1, \sigma_K^2) \,\|\, \mathcal{N}(0, \sigma_K^2)\right),$$
$$\tau(\{\tilde{K}_e^*\}) \triangleq D_\alpha\left((1-\tfrac{b}{N})\mathcal{N}(0, \sigma_K^2) + \tfrac{b}{N}\mathcal{N}(1, \sigma_K^2) \,\|\, \mathcal{N}(0, \sigma_K^2)\right). \tag{19}$$

The total privacy cost of $\{\tilde{K}_{e,1}\}, \{\tilde{K}_e^*\}$ is $\sum_e^E \tau(\{\tilde{K}_{e,1}\}) + \sum_e^E \tau(\{\tilde{K}_e^*\})$.

**Privacy analysis for $\{\tilde{g}_{e,t}\}$ and setting of $\sigma_{G_e}$.** Establishing the privacy cost of $\{\tilde{g}_{e,t}\}$ is challenging since the effect of sampling and weighting in each batch complicate the privacy analysis. We address this challenge with an analysis that borrows ideas from [33]. In what follows, we focus on the privacy cost of one iteration and then compose the costs of all iterations. For simplicity, we denote the clipped gradient $\bar{g}_{e,t}(x_i)$ as $g_i$, $\tilde{K}_{e,t}$ as $\tilde{K}$, and the sampling probability $q_{e,t}(x_i) \cdot p_{e,t}(x_i)$ as $p_i$.

First, by Equation (7), $\tilde{g}$ follows a mixture Gaussian distribution:

$$\mathcal{G}(D) = \frac{1}{b}\sum_{\mathcal{B}} p(\mathcal{B})\mathcal{N}\left(\sum_{i \in \mathcal{B}}\frac{g_i}{\tilde{N}p_i}, \sigma_G^2 C_e^2 \mathbf{I}\right), \tag{20}$$

where $p(\mathcal{B})$ stands for the probability of sampling batch $B$ from dataset $D$ as Equation (8), i.e.,

$$p(\mathcal{B}) = \prod_{x_i}^{\mathcal{B}} p_i \prod_{x_j}^{D\backslash\mathcal{B}}(1 - p_j).$$

Next, consider a neighbor dataset $D' = D \cup \{z\}$. The sets of gradients of $D$ and $D'$ are $\{g_i\}_i^N$ and $\{g_i\}_i^N \cup \{g_z\}$, respectively. Given $D'$ as the training set, the distribution of $\tilde{g}$ is

$$\mathcal{G}(D') = \frac{1}{b} \sum_{\mathcal{B}} p(\mathcal{B}) \Bigg( (1 - p_z) \mathcal{N}\Big( \sum_{i \in \mathcal{B}} \frac{g_i}{\tilde{N} p_i}, \sigma_G^2 C_e^2 \mathbf{I} \Big) +$$
$$p_z \mathcal{N}\Big( \sum_{i \in \mathcal{B}} \frac{g_i}{\tilde{N} p_i} + \frac{g_x}{\tilde{N} p_z}, \sigma_G^2 C_e^2 \mathbf{I} \Big) \Bigg).$$

We proceed to bound the following Rényi divergence:

$$D_\alpha\big( \mathcal{G}(D') \,\|\, \mathcal{G}(D) \big) \leq D_\alpha\big( b \cdot \mathcal{G}(D') \,\|\, b \cdot \mathcal{G}(D) \big)$$
$$\leq \sup_{\mathcal{B}} D_\alpha\Big( (1 - p_z) \mathcal{N}\Big( \sum_{i \in \mathcal{B}} \frac{g_i}{\tilde{N} p_i}, \sigma_G^2 C_e^2 \mathbf{I} \Big)$$
$$+ p_z \mathcal{N}\Big( \sum_{i \in \mathcal{B}} \frac{g_i}{\tilde{N} p_i} + \frac{g_z}{\tilde{N} p_z}, \sigma_G^2 C_e^2 \mathbf{I} \Big) \,\Big\|\, \mathcal{N}\Big( \sum_{i \in \mathcal{B}} \frac{g_i}{\tilde{N} p_i}, \sigma_G^2 C_e^2 \mathbf{I} \Big) \Big)$$
$$\leq \sup_{\|g_z\| \leq C_e} D_\alpha\Big( (1 - p_z) \mathcal{N}(\mathbf{0}, \sigma_G^2 C_e^2 \mathbf{I})$$
$$+ p_z \mathcal{N}\Big( \frac{g_z}{\tilde{N} p_z}, \sigma_G^2 C_e^2 \mathbf{I} \Big) \,\Big\|\, \mathcal{N}\Big( \mathbf{0}, \sigma_G^2 C_e^2 \mathbf{I} \Big) \Big)$$

where first inequality follows from the data processing inequality for Rényi divergence, the second inequality follows from the joint quasi-convex property of Rényi divergence [54], and the last inequality follows from the translation and rotation invariance for Rényi divergence [33]. Since Rényi divergence is additive, for any gradient $g_z$ and its corresponding $p_z$, we have

$$D_\alpha\Big( (1 - p_z) \mathcal{N}(\mathbf{0}, \sigma_G^2 C_e^2 \mathbf{I}) + p_z \mathcal{N}\Big( \frac{g_z}{\tilde{N} p_z}, \sigma_G^2 C_e^2 \mathbf{I} \Big) \,\Big\|\, \mathcal{N}\Big( \mathbf{0}, \sigma_G^2 C_e^2 \mathbf{I} \Big) \Big)$$
$$= D_\alpha\Big( (1 - p_z) \mathcal{N}(0, \sigma_G^2 C_e^2) + p_z \mathcal{N}\Big( \frac{\|g_z\|}{\tilde{N} p_z}, \sigma_G^2 C_e^2 \Big) \,\Big\|\, \mathcal{N}(0, \sigma_G^2 C_e^2) \Big)$$
$$= \frac{1}{\alpha - 1} \ln \left( \sum_{m=0}^{\alpha} \binom{\alpha}{m} (1 - p_z)^{\alpha - m} p_z^m \exp\left( \frac{(m^2 - m)\|g_z\|^2}{\tilde{N}^2 p_z^2 \sigma_G^2 C_e^2} \right) \right).$$

Finally, by Equation (12) and Equation (15), we have

$$\sup_{\|g_z\| \leq C_e} \frac{1}{\alpha - 1} \ln \left( \sum_{m=0}^{\alpha} \binom{\alpha}{m} (1 - p_z)^{\alpha - m} p_z^m \exp\left( \frac{(m^2 - m)\|g_z\|^2}{\tilde{N}^2 p_z^2 \sigma_G^2 C_e^2} \right) \right)$$
$$\leq \frac{1}{\alpha - 1} \ln \left( \sum_{m=0}^{\alpha} \binom{\alpha}{m} (1 - p_z)^{\alpha - m} p_z^m \exp\left( \frac{(m^2 - m)\tilde{K}^2}{\tilde{N}^2 \sigma_G^2 C_e^2} \right) \right)$$
$$\leq \frac{1}{\alpha - 1} \ln \left( \sum_{m=0}^{\alpha} \binom{\alpha}{m} \left( 1 - \frac{bC_e}{\tilde{K}} \right)^{\alpha - m} \left( \frac{bC_e}{\tilde{K}} \right)^m \exp\left( \frac{(m^2 - m)\tilde{K}^2}{2\tilde{N}^2 \sigma_G^2 C_e^2} \right) \right)$$
$$\triangleq \tau(\{\tilde{g}_{e,t}\}).$$

We bridge the total cost of the algorithm from $(\alpha, \tau)$-RDP to $(\epsilon, \delta)$-DP by Equation (2) with

$$\tau = \tau(\tilde{N}) + \sum_e^E \tau(\{\tilde{K}_{e,1}\}) + \sum_e^E \tau(\{\tilde{K}_e^*\}) + \sum_e^E \sum_t^T \tau(\{\tilde{g}_{e,t}\}),$$
$$\text{and } \delta = \delta_0.$$

$\epsilon$ in Equation (2) is an increasing function of $\tau$, where $\tau(\{\tilde{g}_{e,t}\})$ is a decreasing functions of $\sigma_G$ and other terms are constants as $\sigma_N$ and $\sigma_K$ are fixed. Therefore, to ensure $(\epsilon_0, \delta_0)$-DP, we select the smallest $\sigma_G$ that satisfies $\epsilon \leq \epsilon_0$.

**Utility analysis.** We have shown in Section 3 that $\tau(\{\tilde{g}_{e,t}\})$ of DPIS is always equal or less than that of DP-SGD. Moreover, we always have $\sum_e^E \tau(\{\tilde{K}_{e,1}\}) + \sum_e^E \tau(\{\tilde{K}_e^*\}) \ll \sum_e^E \sum_t^T \tau(\{\tilde{g}_{e,t}\})$ in practice because of the gap of their releasing times $2 \cdot E \ll T \cdot E$. Therefore, the total privacy cost of DPIS is less than that of DP-SGD, and hence, we can use a smaller noise multiplier $\sigma_G$ in DPIS and obtain better results.

## 4.2 Privacy Budget Allocation

The last concern of the privacy analysis in DPIS is that we cannot analyze the privacy cost $\tau(\{\tilde{g}_{e,t}\})$ or allocate budget until knowing $\tilde{K}, C$ in every iteration. If DPIS equally allocates the privacy budget per iteration or per epoch like DP-SGD, it may degrade the training efficiency. For instance, as $\tau(\{\tilde{g}_{e,t}\})$ is an increasing function of $\tilde{K}/C$ and a decreasing function of $\sigma_G$, if $\tilde{K}/C$ in later iterations are greater than the incipient ones, we should use a greater $\sigma_G$ to cap the privacy cost. However, the characteristics of the model are on the opposite side. In the early training iterations, the validation accuracy of the model can steadily rise even if a large amount of noise is injected. As the training progresses, the model approaches the bottleneck and becomes more resistant to the injection noise. Therefore, to avoid the increment of $\sigma_G$, we propose the following dynamic budget allocation method:

- We divide the total epochs $E$ as $E = a_E \cdot E + (1 - a_E) \cdot E$ where $a_E$ is a hyperparameter.
- When training in the initial $a_E \cdot E$ epochs we assume that the later $\tilde{K}$ as the worst case, i.e., $\tilde{K} = \tilde{N}C$.
- When training in the last $(1 - a_E) \cdot E$ epochs we equally allocate the remaining budget, i.e., all $\tilde{K}$ in future equals the current one.
- For the clipping threshold $C$, we assume that it is invariant in future whenever we perform the privacy analysis.

In our implementation, we divide the training epochs into two phases. During the first phase (i.e., initial training), we overestimate the privacy budget required by later epochs, by assuming that all future epochs use the worse-case value of $\tilde{K}$, i.e., $\tilde{N}C$. Then with sufficient budget, later $\sigma_G$ will not increase as the training goes. In the second phase (i.e., closer to convergence), we assume that all future values of $\tilde{K}$ are identical to its value in the current epoch, which is clearly smaller than the worst-case value. This phase equally allocates the budget to the remaining epochs and leaves more iterations for updating the parameters in a low injection noise environment. As shown in Figure 1, the time when noise drops off is exactly when the training enters the second phase. How much the noise drops depends upon how much smaller the real $\tilde{K}$ is compared to its worst-case value $\tilde{N}C$. In experiments with raw pixels and texts as input, the noise drops by 40-50% on MNIST and FMNIST, by 10-20% on CIFAR-10 and by 20-30% on IMDb, respectively.

We have established the allocation principle of each epoch so far. For each iteration in the same epoch, we use the same $\tilde{K}$ defined in Equation (12).

# 5 EXPERIMENTS

This section empirically evaluates DPIS against four competitors. All algorithms are implemented in Python using Tensorflow (www.tensorflow.org), following the same framework as the Tensorflow Privacy library (github.com/tensorflow/privacy).

## 5.1 Competitors

We compare DPIS against DP-SGD [2] and three state-of-the-art solutions built upon it: DP-SGD with handcrafted features [52], DP-SGD with tempered sigmoid activation [42], and DP-SGD with exponential decay noise [64], which we refer to as DP-SGD-HF, DP-SGD-TS, and DP-SGD-Decay, respectively.

Note that we do not consider other approaches [9, 13, 37, 39] that are orthogonal to DPIS. For example, we do not include the method in [13] as it tackles the modification of over-parameterized model structures and the tuning of hyper-parameters, which considerably differs from our focus. As another example, the semi-supervised model PATE [39] requires an additional public dataset, which is unavailable in our problem setting.

## 5.2 Datasets and Models

We use four common benchmarks for differentially private deep learning: MNIST, Fashion MNIST (FMNIST), CIFAR-10, and IMDb. We construct multilayer neural networks for these benchmarks with Keras [10]. Unless otherwise specified, we let each layer follow the default setting in Keras.

**MNIST [25].** This is an image dataset of handwritten digits from 10 categories (7,000 images per category). It contains a training set of 60,000 examples and a test set of 10,000 examples. Each example contains a $28 \times 28$ grayscale image and a label denoting its category. In the non-private setting with handcrafted features as inputs, the model reaches 99.1% accuracy in 20 epochs [52].

**FMNIST [62].** This is an image dataset of fashion products from 10 categories (7,000 images per category). It contains a training set of 60,000 examples and a test set of 10,000 examples. Each example contains a $28 \times 28$ grayscale image and a label denoting its category. In the non-private setting with handcrafted features as inputs [52], the model reaches 90.9% accuracy in 20 epochs.

**CIFAR-10 [24].** This is an image dataset of colored objects from 10 categories (6,000 images per category). It contains 50,000 training examples and 10,000 testing examples. Each example contains a colour image of resolution $32 \times 32$ with three colour channels, and a label denoting its category. In the non-private setting with handcrafted features as inputs [52], the model reaches 71.0% accuracy in 20 epochs.

For DPIS on MNIST, FMNIST, and CIFAR-10, we follow the preprocessing steps described in [52] and use the same convolutional neural network architecture as DP-SGD-HF. For DP-SGD-TS, DP-SGD-Decay, and DP-SGD, we use the convolutional neural network architecture described in [42]. The loss function for all methods are categorical cross-entropy.

**IMDb [28].** This dataset from the Internet Movie Database contains 50,000 movies reviews with obvious bias (positive or negative),

**Table 1: RNN model for IMDb.**

| Layer | Parameters |
|---|---|
| Embedding | 100 units |
| Fully connected | 32 units |
| Bidirectional LSTM | 32 units |
| Fully connected | 16 units |
| Fully connected | 2 units |

**Table 2: Setting of parameters in DPIS**

| Parameter | MNIST | FMNIST | CIFAR-10 | IMDb |
|---|---|---|---|---|
| $k$ | 5 | 5 | 5 | 3 |
| $\sigma_K$ | \multicolumn{4}{c}{$0.02 \cdot N$} |
| $a_E$ | 0.8 | 1 | 0.8 | 0.8 |
| $C^*$ | \ | \ | \ | $4 \cdot C_1$ |
| $\lambda$ | \ | \ | \ | 1 |

among which 25,000 are in the training set and 25,000 are in the test set. Each review has been encoded as a list of word indexes. Words are indexed by their overall frequency in the dataset, and we only consider the top 20,000 most common words. We pad all sequences to 80 words long to ensure that all inputs have the same shape. We use a five-layer recurrent neural network shown in Table 1, where the embedding layer is loaded with parameters from the GloVe word embedding [43]. For the second and fourth layers, we use Relu activation in DP-SGD, DPIS, and DP-SGD-Decay, and use tanh activation in DP-SGD-TS. We freeze the parameters in the embedding layer and only train the other four layers. The loss function is cross-entropy. In the non-private setting with an expected batch size of 32, Adam optimizer and other default parameters, the model achieves 79.5% accuracy in 20 epochs.

## 5.3 Parameter Settings

In our experiments, we vary the privacy budget $\epsilon \in \{0.5, 1, 2, 3, 4\}$ for every dataset, while fixing $\delta$ to a small value $10^{-5}$. In order to fairly compare DPIS with other competitors, we set an identical $\sigma_N = 0.02 \cdot N$ for them, so that their privacy budget allocated to protect $N$ and the training process are equal. We fix the number of iterations in each epoch to $T = |D|/b$ instead of $\hat{N}/b$, to avoid the difference in the number of iterations caused by $\hat{N}$ in experiments.

We use two SGD optimizers: SGD with momentum [10] and Adam [10]. For the former, we set the momentum to 0.9; for the latter, we set the exponential decay factors as $\beta_1 = 0.9$ and $\beta_2 = 0.999$. We follow the best settings of the expected batch size $b$, learning rate $\eta$, the initial clipping bound $C_1$, and parameters in the scatter net tuned by [52]. In addition, we tune the number of training epochs $E$ based on the corresponding privacy budget. Following common practice in the literature (e.g., explained in [52]), we do not account for the privacy loss of this hyperparameter tuning step, since such tuning is common in all differentially private machine learning methods.

DPIS has a number of parameters: the sampling probability multiplier $k$, the privacy budget allocator $a_E$, the noise multiplier $\sigma_K$, the external clipping threshold $C^*$, and the quantile $\lambda$. Table 2 shows our setting of these parameters. Note that $C^*$ and $\lambda$ are irrelevant on MNIST, FMNIST, and CIFAR-10, because we do not apply adaptive clipping on these datasets except for the ablation study (see Section 5.5). For DP-SGD-Decay, we choose the decay rate in [0.001, 0.01] according to [64].

## 5.4 Model Accuracy Comparison

Table 3 shows the classification accuracy of each method, averaged over 5 runs, with $\epsilon$ varying in {0.5, 1, 2, 3, 4} (We omit DP-SGD-HF on the IMDb dataset because its scattering network [52] cannot be applied to natural language processing.) Observe that DPIS consistently outperforms the competitors in all settings. In particular, on IMDb, its classification accuracy is 2-3% higher than the best of the competing methods. In comparison, on MNIST, FMNIST, and CIFAR-10, the performance gap between DPIS and the competitors is not as substantial, because the accuracy of [52] is already close to that obtained in the non-private setting, *i.e.*, there is little room for further performance improvement.

The reason that *DPIS* significantly outperforms competing methods on IMDb is as follows. First, the neural network trained on IMDb is an RNN, and the $L_2$ norm and variance of gradients on this RNN are much larger than those on the CNNs used for MNIST, FMNIST, and CIFAR-10. Consequently, our importance sampling approach is more effective on variance reduction and noise reduction on IMDb. Second, as our clipping bound increases with the gradients through adaptively clipping, the amplitude of the noise (*i.e.*, $\sigma_G^2 \cdot C_\epsilon^2$) decreases more significantly when DPIS reduces $\sigma_G$ during training.

Observe that DPIS achieves roughly the same accuracy as other competitors when it uses only half of their privacy budget on MNIST, FMNIST, and IMDb. In other words, DPIS provides the same model utility with twice as much strength of privacy guarantee as other methods. It is also worth noting that DP-SGD-Decay, whose idea is to allocate more privacy budget to later epochs of training, does not ensure higher model utility than DP-SGD-TS and DP-SGD, which allocate the budget to each epoch equally. In contrast, DPIS is able to improve the effectiveness of later epochs due to its advanced sampling scheme.

Summarizing the comparison results, DPIS demonstrates significant and consistent performance advantages in terms of accuracy, compared to the state-of-the-art. Further, its utility does not degrades significantly as the privacy budget decreases. Hence, it is the method of choice in practical applications.

## 5.5 Ablation Study

Table 4 compares the accuracy of DPIS with and without adaptive clipping, referred to as DPIS+AC and DPIS\AC, respectively. Observe that the adaptive gradient clipping optimization noticeably improves the performance of DPIS on IMDb, but is less effective on MNIST, FMNIST, and CIFAR-10. The reason is that, on IMDb, the norms of gradients vary significantly (in the range of [3, 10]) during training, and hence, there is much benefit in adaptively adjusting the clipping bound. In contrast, on MNIST, FMNIST, and CIFAR-10, the norms of gradients remain small throughout the training process, which leaves little room for adaptive clipping to improve the accuracy of DPIS.

Tables 5 and 6 show the accuracy of DPIS with varying budget scheduler $a_E$ and probability multiplier $k$, respectively. We observe from Table 5 that $a_E = 0.8$ leads to favorable results on all datasets. This observation is consistent with the motivation of our privacy budget schedule: it is beneficial to save some privacy budget of DPIS in early-stage training, and spend it when the training approaches the bottleneck in the later stage. Meanwhile, Table 6 shows that the performance DPIS is insensitive to the choice of $k$ in the range of [3, 6]. We have also evaluated the effects of two other parameters of DPIS: the external clipping threshold $C^*$ and the quantile $\lambda$; we include the results in [59].

## 6 RELATED WORK

There exists a plethora of techniques for differentially private data analysis in the literature [17, 67]. The design of existing techniques often consists of the following steps: (i) identifying the function to release, (ii) bounding the sensitivity of the function, and (iii) designing a suitable noise injection method based on the sensitivity. However, it is difficult to apply the same methodology to privacy preserving deep learning. The challenges lie in the complicated structure of the neural networks, the non-convexity of the objective function (which makes it difficult to analyze sensitivities), and the large number of iterations required for training (which leads to difficulties in reducing privacy loss).

Abadi *et al.* [2] are the first to introduce a viable method, DP-SGD, for training deep learning models with differential privacy. The moments accountant (MA) approach in [2] has also been applied to address several other problems, *e.g.*, empirical risk minimization [27, 56] and machine learning as a service [26]. In 2019, Mironov [33] proposes a new analysis approach for precise computation of privacy based on RDP, which outperforms the moments accountant and greatly improved the performance of DP-SGD. Subsequent work [9, 37, 42, 52] shifts attention to slightly modify the model structure or the learning algorithm. The state-of-the-art [42] replaces the ordinary activation functions like ReLU in CNNs with tempered sigmoid functions. Another branch of work [44, 61, 64] focuses on optimizations such as adaptively adjusting the gradient clipping bound or adaptively allocating the privacy budget. But it is not trivial to embed importance sampling into their clipping methods and these methods are still inferior to DPIS with our adaptive clipping algorithm in terms of the accuracy of the model trained. Besides, the privacy allocation mechanism [64] can be directly applied to DPIS without other changes.

Both the composition and privacy amplification via subsampling in DP-SGD and our analysis are well studied in the literature. Tight bounds and constructive algorithms for composition are studied in [22, 30]. Privacy amplification via subsampling is studied in [3, 58, 68]. Our privacy analysis borrows ideas from [33], which quantifies the privacy loss of the Gaussian mechanism when subsampling is used. The main difference between our analysis and that in [33] is that we modify the sampling probability and the mean of the Gaussian distribution and transfer the difference between two Gaussian mixture distributions from the mean to the outside probability as shown in Equation (21).

While we analyze $(\epsilon, \delta)$-DP for the privacy guarantee of DPIS, there also exists several other variants of DP, *e.g.*, Concentrated-DP [18], zero Concentrated-DP [6], truncated Concentrated-DP [5],

**Table 3: Algorithm Comparison: Testing Accuracy.**

| Dataset | Method | $\epsilon = 0.5$ | $\epsilon = 1$ | $\epsilon = 2$ | $\epsilon = 3$ | $\epsilon = 4$ |
|---------|--------|------------------|----------------|----------------|----------------|----------------|
| MNIST | DPIS | **96.8%** | **97.7%** | **98.6%** | **98.8%** | **98.9%** |
| (non-DP 99.1%) | DP-SGD-HF [52] | 96.1% | 97.2% | 98.3% | 98.4% | 98.5% |
| | DP-SGD-TS [42] | 96.0% | 96.8% | 97.7% | 98.1% | 98.3% |
| | DP-SGD-Decay [64] | 96.2% | 97.0% | 97.8% | 98.2% | 98.3% |
| | DP-SGD [2] | 93.1% | 94.9% | 96.1% | 96.8% | 97.2% |
| FMNIST | DPIS | **84.6%** | **86.6%** | **88.3%** | **88.8%** | **89.4%** |
| (non-DP 90.9%) | DP-SGD-HF [52] | 83.9% | 86.0% | 87.8% | 88.3% | 88.7% |
| | DP-SGD-TS [42] | 81.0% | 83.0% | 85.0% | 86.3% | 86.8% |
| | DP-SGD-Decay [64] | 80.8% | 83.1% | 85.1% | 86.4% | 86.8% |
| | DP-SGD [2] | 78.4% | 80.8% | 82.3% | 84.1% | 84.5% |
| CIFAR-10 | DPIS | **55.1%** | **62.2%** | **67.9%** | **69.7%** | **70.6%** |
| (non-DP 71.0%) | DP-SGD-HF [52] | 54.3% | 61.1% | 67.4% | 69.2% | 70.2% |
| | DP-SGD-TS [42] | 44.4% | 51.0% | 57.2% | 60.7% | 63.0% |
| | DP-SGD-Decay [64] | 44.1% | 50.7% | 56.6% | 60.7% | 63.1% |
| | DP-SGD [2] | 43.1% | 48.3% | 54.7% | 58.9% | 60.4% |
| IMDb | DPIS | **59.1%** | **62.1%** | **65.9%** | **68.3%** | **70.2%** |
| (non-DP 79.5%) | DP-SGD-TS [42] | 57.5% | 60.3% | 62.8% | 65.2% | 66.5% |
| | DP-SGD-Decay [64] | 56.8% | 60.4% | 63.6% | 66.3% | 67.6% |
| | DP-SGD [2] | 56.4% | 60.3% | 63.5% | 66.4% | 67.4% |

**Table 4: Comparing DPIS+AC and DPIS\AC on testing accuracy.**

| Dataset | Method | $\epsilon = 0.5$ | $\epsilon = 1$ | $\epsilon = 2$ | $\epsilon = 3$ |
|---------|--------|------------------|----------------|----------------|----------------|
| MNIST | DPIS+AC | 96.7% | 97.6% | 98.5% | 98.7% |
| | DPIS\AC | **96.8%** | **97.7%** | **98.6%** | **98.8%** |
| FMNIST | DPIS+AC | **84.6%** | 86.5% | 88.2% | 88.7% |
| | DPIS\AC | **84.6%** | **86.6%** | **88.3%** | **88.8%** |
| CIFAR-10 | DPIS+AC | **55.4%** | **62.4%** | 67.7% | 69.3% |
| | DPIS\AC | 55.1% | 62.2% | **67.9%** | **69.7%** |
| IMDb | DPIS+AC | **59.1%** | **62.2%** | **65.9%** | **68.3%** |
| | DPIS\AC | 58.2% | 61.7% | 65.0% | 67.6% |

**Table 5: Performance of DPIS with varying privacy budget scheduler $a_E$.**

| Dataset | $a_E = 0$ | $a_E = 0.6$ | $a_E = 0.8$ | $a_E = 1$ |
|---------|-----------|-------------|-------------|-----------|
| MNIST | 97.6% | 97.6% | 97.7% | 97.6% |
| FMNIST | 86.6% | 86.5% | 86.7% | 86.8% |
| CIFAR-10 | 62.0% | 62.1% | 62.4% | 61.9% |
| IMDb | 62.1% | 61.8% | 62.1% | 62.1% |

**Table 6: Performance of DPIS with varying probability multiplier $k$.**

| Dataset | $k = 3$ | $k = 4$ | $k = 5$ | $k = 6$ |
|---------|---------|---------|---------|---------|
| MNIST | 97.7% | 97.7% | 97.7% | 97.7% |
| FMNIST | 86.7% | 86.7% | 86.8% | 86.8% |
| CIFAR-10 | 62.3% | 62.2% | 62.4% | 62.4% |
| IMDb | 62.1% | 62.2% | 62.3% | 62.3% |

## 7 CONCLUSION

This paper presents a novel solution, DPIS, for differentially private neural network training. Through a stage-setting sampling technique, adaptive mechanisms for tuning parameters in a differentially private manner, as well as sophisticated privacy loss analysis following the RDP framework, DPIS achieves significant and consistent accuracy gains over the existing methods. We formally prove the correctness of DPIS, and conduct extensive experiments on classic benchmark datasets to evaluate the effectiveness of DPIS on different deep models. The results demonstrate high accuracy obtained by DPIS, which may become the enabling technology for deep learning with differential privacy. For future work, we plan to accelerate the training process of DPIS and expand the sampling pool to improve its performance.

## ACKNOWLEDGMENTS

This work was supported by the Ministry of Education, Singapore (Number MOE2018-T2-2-091), A*STAR, Singapore (Number A19E3b0099), and Qatar National Research Fund, a member of Qatar Foundation (Number NPRP11C-1229-170007). Any opinions, findings and conclusions or recommendations expressed in this material are those of the authors and do not reflect the views of the funding agencies.

and Rényi-DP [32]. These variants are designed for different settings, and could be translated into another under certain conditions. We focus on $(\epsilon, \delta)$-DP since it is the most commonly used in the literature and in practice. An orthogonal line of work studies privacy preserving machine learning under the decentralized setting, e.g., [29, 39, 40, 47, 57]. Another complimentary line of work considers how to extract sensitive information in machine learning, e.g., [34, 35, 49].